\documentclass[a4paper]{elsarticle}

\pdfoutput=1
\usepackage[colorlinks,bookmarks]{hyperref}

\usepackage[T1]{fontenc}
\usepackage[utf8]{inputenc}
\usepackage{lmodern}

\usepackage{amsmath}

\newcommand{\be}{\begin{equation}}
\newcommand{\ee}{\end{equation}}
\newcommand{\bea}{\begin{eqnarray}}
\newcommand{\eea}{\end{eqnarray}}
\newcommand{\ba}{\begin{eqnarray}}
\newcommand{\ea}{\end{eqnarray}}

\def\d {\mathrm{d}}

\newcommand{\aperp}{a_{\perp }}
\newcommand{\aperpo}{a_{\perp_0}}
\newcommand{\apar}{a_{\parallel}}

\newcommand{\Hperp}{H_{\perp}}
\newcommand{\Hperpo}{H_{\perp_0}}
\newcommand{\Hpar}{H_{\parallel}}

\begin{document}

\begin{frontmatter}
\title{Intrinsic uncertainty on the nature of dark energy}

\author{Wessel Valkenburg}
\address{Instituut-Lorentz for Theoretical Physics, Universiteit Leiden, Postbus 9506, 2333 CA Leiden, The Netherlands}

\author{Martin Kunz}
\address{D\'epartment de Physique Th\'eorique and Center for Astroparticle Physics, Universit\'e de Gen\`eve,
Quai E. Ansermet 24, CH-1211 Gen\`eve 4, Switzerland}
\address{African Institute for Mathematical Sciences, 6-8 Melrose Road, Muizenberg, Cape Town, South Africa}

\author{Valerio Marra}
\address{Institut f\"ur Theoretische Physik, Universit\"at Heidelberg, Philosophenweg
16, 69120 Heidelberg, Germany}

\begin{abstract}
We argue that there is an intrinsic noise on measurements of the equation of state parameter $w=p/\rho$ from large-scale structure around us. The presence of the large-scale structure leads to an ambiguity in the definition of the background universe and thus there is a maximal precision with which we can determine the equation of state of dark energy. To study the uncertainty due to local structure, we model density perturbations stemming from a standard inflationary power spectrum by means of the exact Lema\^{i}tre-Tolman-Bondi solution of Einstein's equation, and show that the usual distribution of matter inhomogeneities in a $\Lambda$CDM cosmology causes a variation of $w$ -- as inferred from distance measures -- of several percent.
As we observe only one universe, or equivalently because of the cosmic variance, this uncertainty is systematic in nature.
\end{abstract}

\end{frontmatter}

\section{Introduction}

A key quantity to characterize the nature of the dark energy is its equation of state parameter
$w=p/\rho$. Current and future cosmological observations try to measure $w$ ever more
accurately, and the power of dark-energy missions is judged by the minimal error that
they can achieve on the dark-energy equation of state parameter $w=p/\rho$. This is for example
the basis of the Dark Energy Task Force (DETF) \cite{Albrecht:2006um} Figure of Merit (FoM), which is given by the
determinant of the Fisher matrix for the parameters $w_0$ and $w_a$ in a linear parameterization
of the equation of state, $w(a) = w_0 + (1-a) w_a$. An important question in the context of 
dark-energy research is then whether there is a natural limit for the precision with which
$w$ can be measured, or whether one can in principle determine $w(a)$ to an arbitrary precision.

In this paper we argue that the matter fluctuations that are always present in the universe provide
such a limit, and we determine the unavoidable variation of $w(a)$ as expected in the $\Lambda$CDM concordance model
(for which in principle $w=-1$). Those variations appear because we always measure
{\em any} observable quantity in the true perturbed universe, even if we consider ``background'' quantities like the luminosity distance (see e.g.~\cite{Bonvin:2005ps} and references therein), and they remain
significant even when averaging over angles \cite{Mustapha:1997xb}. This is a direct
manifestation of the ``fitting problem''~\cite{Ellis:1987zz}, i.e.\ the attempt to fit a homogenous and isotropic FLRW model to a lumpy universe~\cite{Marra:2007gc, Kolb:2009rp,Clifton:2009nv,Bull:2012zx} rather than directly modeling the inhomogeneities~\cite{Neill:2007fh,Li:2007ny,Sinclair:2010sb,Romano:2010nc,Valkenburg:2011ty,Wiegand:2011je,Romano:2011mx,deLavallaz:2011tj,Valkenburg:2012ds,Marra:2012pj,BenDayan:2012ct,BenDayan:2013gc,Marra:2013rba}.
If, on one hand, fluctuations in e.g.\ the luminosity distance allow us in principle to obtain additional cosmological information (see e.g.\ \cite{Valageas:1999ch,Dodelson:2005zt,Sarkar:2007sp,Bonvin:2006en, Amendola:2010ub,Amendola:2013twa,Quartin:2013moa}),
on the other hand they result in an intrinsic noise in the determination of cosmological parameters.
Indeed, although the perturbations in the metric are small, only about $10^{-5}$, they can be amplified when going to quantities that involve derivatives like $w(a)$, as demonstrated in e.g.~\cite{Clarkson:2007bc}.

The key ingredient that is new in this analysis, is a quantification of the level of ignorance about the position of us as the observer. Our local gravitational potential relative to cosmological scales is unknown to us. We do know which level of variance the $\Lambda$CDM paradigm predicts for our potential. It is this variance that consequently implies a systematic error in the determination of the homogeneous universe's $w(a)$, the equation of state of Dark Energy.

The ignorance could be alleviated if observations were able to constrain the density field around the observer. However, currently this is far from possible. Current observations are, even in the most ambitious analysis including all possible distance measures as well as Compton-y distortion, the kinematic Sunyaev-Zel'dovich effect and ages of galaxies versus redshift, not capable of constraining the matter distribution in our local patch of the universe down to the level of density perturbations expected in the standard inflationary scenario~\cite{Valkenburg:2012td}.

The outline of this paper is as follows: first we define the probability of inhomogeneities. Then we explain how we model
said inhomogeneities. Next we show the noise caused by this structure on the evolution of the dark-energy equation of state inferred by an observer who ignores the inhomogeneity.
Finally we obtain the error in the cosmological parameters measured by an observer who fits the luminosity distance under various assumptions.

\section{Probability of a local structure}

The root mean square of the density perturbation in a sphere of radius $L$ around any point today in a Gaussian density field is~\cite{Kolb:1990vq}
\begin{align} \label{sila}
\sigma_L = \left[  \int_0^\infty dk \frac{k^2}{2\pi^2}P(k)  \left(\frac{3j_1 (Lk)}{Lk}\right)^2   \right]^{\tfrac{1}{2}},
\end{align}
where $P(k)$ is the matter power spectrum today 
as a function of wavenumber $k$, and $j_1$ is the spherical Bessel function of the first kind.

Given some inhomogeneity with a mass $M(L)$ inside a radius $L$,  one can define the average density perturbation 
$\delta_0 \equiv M(L) / \bar M(L) -1$ relative to the homogeneous background which predicts a mass $\bar M(L)$ inside the same radius $L$.  Then the probability of having such a structure is~\cite{Valkenburg:2012td},
\begin{align} \label{gauprob}
P(\delta_0|L) = \frac{1}{\sigma_{L} \sqrt{2 \pi}} \; e^{-\tfrac{\delta_{0}^{2}}{2 \, \sigma_{L}^{2}}} \,.
\end{align}

Since the study of Dark Energy entails mainly a search for
a redshift dependent effect, it is a search for a radially dependent effect in a spherical coordinate system with the observer at the origin. The observer hence averages over all angles, which is equivalent to expanding the full matter field in spherical harmonics, and throwing away all other information than the radially dependent monopole.
This should be a reasonable approximation as long as distance measurements are averaged over angles in
analyses of the dark energy equation of state. 
Hence we model the inhomogeneity spherically symmetric
and the mass is given by $M(r)\equiv 4\pi \int_0^r dr \sqrt{-g} \rho_{m}(r)$ with $g$ the determinant of the metric.
\section{Model for local inhomogeneity}

To model the inhomogeneity averaged over angular directions, we adopt the spherically symmetric Lema\^{i}tre-Tolman-Bondi solution~\cite{Lemaitre:1933gd, Tolman:1934za, Bondi:1947av} including a cosmological constant $\Lambda$ ($\Lambda$LTB, see e.g.~\cite{Sinclair:2010sb, Marra:2010pg,Valkenburg:2011tm, Romano:2011mx}), for which we can compute all distance measures exactly.
The use of the exact LTB model allows us to deal with nonlinear inhomogeneities, which will be encountered at small radii or low redshifts (their contrast is of the order of $\sigma_L$). Structures of larger radii could have been equally well modeled using linear theory.

The $\Lambda$LTB metric in the comoving and synchronous gauge can be written as (using units for which $c=1$)
\ba \label{metric}
\d s^2 = -\d t^2 + \frac{\apar^2(t,r)}{1-k(r)r^2}\d r^2 + \aperp^2(t,r)r^2 \, \d\Omega^2 \,,
\ea
where the longitudinal ($\apar$) and perpendicular ($\aperp$) scale factors are related by
$\apar = (\aperp r)'$, and a prime denotes partial derivation with respect to the coordinate radius $r$.
In the limit $k\to$\,const., and $a_\perp=a_\|$ we recover the FLRW metric, but in a LTB metric
the curvature $k(r)$ is a free function and in general is not constant.

The two scale factors define two different Hubble rates:
\ba
\Hperp(t,r) \equiv \frac{\dot a_\perp}{\aperp}\,,~~~~~~~~~~~~~
\Hpar(t,r) \equiv \frac{\dot a_{\|}}{\apar} \,.
\ea
The analogue of the Friedmann equation in this space-time can be written in a familiar form,
\be \label{frieq}
\frac{\Hperp^2}{\Hperpo^2}=\Omega_m \, \aperp^{-3} + \Omega_k \, \aperp^{-2} + \Omega_\Lambda \,,
\ee
where we adopted the gauge fixing $\aperpo=1$.
However, the density parameters are now also functions of $r$,
\be \label{omegas}
\Omega_m(r)=\frac{m(r)}{\Hperpo^2 } ,
~~~
\Omega_k(r)=-\frac{k}{\Hperpo^2},
~~~
\Omega_\Lambda(r)=\frac{\Lambda}{3\Hperpo^2 } ,
\ee
so as to satisfy $\Omega_m(r)+ \Omega_k(r)+\Omega_\Lambda(r)=1$.
The free function $m(r)$ is related to the local matter density
$8 \pi G \, \rho_{m}(t,r) = (m  r^{3})' / a_{\parallel}a_{\perp}^2 r^2$.

Finally, time $t$ and radius $r$ as a function of redshift $z$ are determined on the past light cone of the central observer by the differential equations for radial null geodesics,
\ba
\frac{\d t}{\d z} = -\frac{1}{(1+z)\Hpar}\,, ~~~~~~
\frac{\d r}{\d z} = \frac{\sqrt{1-k r^2}}{(1+z)\apar\Hpar} \, , \label{eq:geodesics}
\ea
with the initial conditions $t(0) = t_0$ and $r(0)=0$.
The area ($d_A$) and luminosity ($d_L$) distances are given~by
\be
d_A(z)=a_\perp \big(t(z),r(z) \big) \; r(z), ~~~~ d_L=(1+z)^2 d_A \,.  \label{eq:ltbdist}
\ee

\section{Density profile}

The age of the universe is  a function of $(t,r)$ and is obtained by integrating the Friedmann equation (\ref{frieq}) from the big-bang time $t_{\rm bb}(r)$ to time $t$:
\be \label{tbb}
t - t_{\rm bb} = \frac{1}{\Hperpo(r)}\int_{0}^{\aperp (t,r)} \!\!\!\!\!\!\!\!\!\!\!\!\! \frac{\d  x}{\sqrt{\Omega_m (r)x^{-1} + \Omega_k (r) +  \Omega_\Lambda(r) x^{2}}} .
\ee
Eq.~(\ref{tbb}) relates the three free functions $t_{\rm bb}(r)$, $k(r)$ and $m(r)$, so that density of the dust field in the $\Lambda$LTB model is specified by two free functional degrees of freedom, where we choose $k(r)$ and $t_{\rm bb}(r)$.
Any radial dependence of $t_{\rm bb}(r)$ is directly related to a decaying mode in the matter density field~\cite{1977A&A....59...53S,Zibin:2008vj}. By choosing $t_{\rm bb}(r)=0$ decaying modes are absent, in agreement with the standard scenario of inflation.

We parameterize  the curvature function with the monotonic profile
\begin{align} \label{profi1}
k(r)= k_{b} + (k_c - k_{b}) \; P_3 (r/ r_b ) \,,
\end{align}
where $r_b$ is the comoving radius of the spherical inhomogeneity and
$P_3$ is the function
\begin{align} \label{Pnf}
P_{n}(x)= \left\{\begin{array}{ll}
1-e^{- \left(1-x \right)^n/x } & \mbox{ for }  0  \le x < 1\\
0 & \mbox{ for } \phantom{0. \le} x \geq 1
\end{array}\right. \nonumber
\end{align}
for $n=3$.  The function $P_n(x)$ interpolates from $1$ to $0$
when $x$ varies from $0$ to $1$ while remaining $n$ times differentiable, which implies that
that $k(r)$ is $C^n$ 
everywhere. We choose $n=3$, such that the metric is $C^2$ and the Riemann curvature is $C^0$. For $r \ge r_b$ the curvature profile equals the curvature $k_{b}$ of the background such that for $r \ge r_b$ the metric reduces exactly to the $\Lambda$CDM model. The central under- or over-density, determined by the curvature $k_c$ at the center, is automatically compensated by a surrounding over- or under-dense shell.
We adopt the conservative approach of using a compensated density profile so as not to alter the background metric of the universe, which otherwise would be FLRW only asymptotically.
The radius $L$ of the inhomogeneity that is used in Eq.~\eqref{gauprob} is the radius at which the central over- or under-density has the transition to the surrounding compensating under- or over-dense shell, that is the radius at which the contrast goes through zero.
The radius $L$ is hence smaller than the radius $r_b$ which defines the radius of the total LTB patch, including both the central perturbation and its compensating shell.
The definition of $L$ we have adopted unambiguously and naturally defines the central structure, because it includes all the overdense (underdense) central density perturbation, and excludes all the compensating structure.

In summary, the local structure is parametrized by the radius of the boundary $r_{b}$ and the central curvature $k_{c}$.
For any choice of these parameters, and for a $\Lambda$CDM cosmology (given by the background matter density, the curvature parameter  $k_{b}$, the present-day background Hubble rate and a specific $P(k)$), we can compute the probability of the existence of such a structure using Eq.~\eqref{gauprob} and compute the luminosity distance from an object to the central observer as a function of redshift using Eqs.~(\ref{eq:geodesics},\ref{eq:ltbdist}). Note that for Eq.~\eqref{gauprob} we could have used the smooth density profile from Eq.~\eqref{profi1}, but the difference on the analysis turns out to be negligible.

\section{FLRW Observer's \boldmath{$w(z)$}}

Following \cite{Clarkson:2007bc} one can -- given a luminosity distance-redshift relation in a homogeneous universe described by the FLRW metric -- compute what the underlying $w(z)$ of the dark-energy fluid is. There is  an exact relation between $w(z)$ and the first and second derivatives of the luminosity distance with respect to redshift and two more parameters, $\Omega_k$ and $\Omega_{m}$.
Therefore, the observer can derive $w(z)$ if he/she knows the latter two parameters from other observations, and deduces the derivatives of the luminosity distance from SN observations.
In the scenario studied here, $w=-1$. However, the inhomogeneities come into play and modify the luminosity distance-redshift relation. Consequently, an observer that erroneously assumes that the metric surrounding him/her is FLRW will in fact see a redshift-dependent $w$~\cite{Shafieloo:2009ti, Marra:2012pj, Zhao:2012aw, BenDayan:2012ct}.

In order to study this fundamental variation of the reconstructed $w(z)$, we first set the fiducial flat $\Lambda$CDM model to the WMAP7+LRG best-fit cosmology~\cite{Komatsu:2010fb}.
We then sample the $\{r_{b}, k_{c} \}$ parameter space by building a Markov chain using Eq.~(\ref{gauprob}) as likelihood, but restricted to a certain range in redshifts $0.1(i-1) < z_b \le 0.1i$ for $i=1..12$,\footnote{The expression $1..12$ is short for $1,...,12$.} where $z_b=z(r_b)$ is the apparent redshift at which an observer sees the radius $r_b$ (see Eq.~\eqref{eq:geodesics}). Next we compute from Eq.~(\ref{eq:ltbdist}) the corresponding luminosity distances in the various realizations. We finally derive using Eq.~(3) of \cite{Clarkson:2007bc} the (apparent) $w(z)$ that an FLRW observer would infer. In this procedure we let the FLRW observer fix $\Omega_m$ to the fiducial value as it cannot be determined from cosmological observations alone when allowing for an arbitrary equation of state of the dark energy \cite{Kunz:2007rk}. The curvature could in principle be constrained by combining measurements of the distances with measurements of the expansion rate $H(z)$, but for simplicity we also fix $\Omega_k$ to the fiducial (zero) value.
This simplifies Eq.~(3) of \cite{Clarkson:2007bc} to:
\begin{equation}\label{weq}
w(z)=\frac{\frac{2}{3}(1+z)[(1+z){D}_{L}'-{D}_{ L}]^{-1}\left\{[(1+z)^2]{D}_{
L}''-\frac{1}{2}[(1+z){D}_{ L}'-{D}_{ L}]\right\}}
{(1+z)[\Omega_{ m}(1+z)]{D}_{L}'^2-2[\Omega_{ m} (1+z)] {D}_{ L}{D}_{
L}'+\Omega_{ m} {D}_{ L}^2-(1+z)} ,
\end{equation}
where $'=\mathrm{d}/\mathrm{d}z$ and ${D}_L=(H_0/c) d_L$ is the dimensionless luminosity distance.

\begin{figure}
\includegraphics[width=\textwidth]{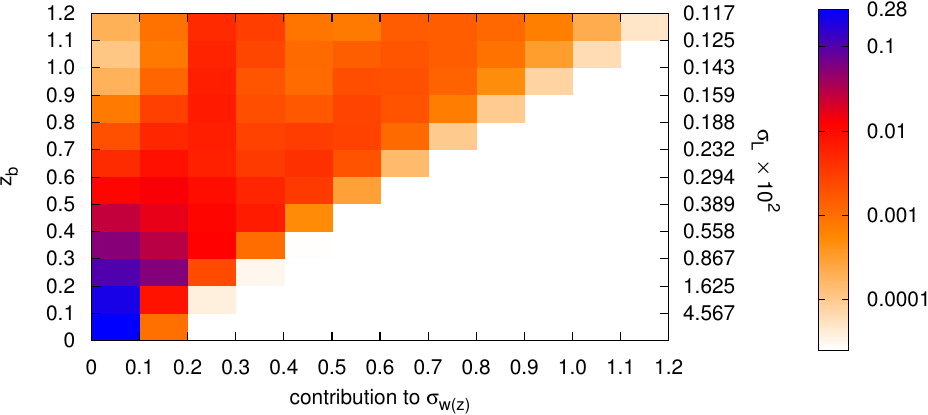}
\caption{Matrix representation of the contribution of matter perturbations with a radius $0.1(i-1) < z_b \le 0.1i$ (the index $i=1..12$ labels rows) to the dispersion $\sigma_{w(z)}$ at $z_j=0.05 + 0.1 j$ (the index $j=1..12$ labels columns). The right vertical axis shows the variance of matter perturbations inside the corresponding radius on the left vertical axis.}
\label{fig:matrix}
\end{figure}

\begin{figure}
\includegraphics[width=\columnwidth]{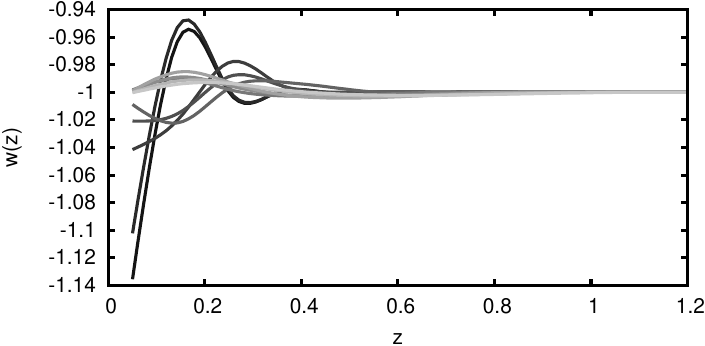}
\caption{Examples of apparent evolution of the dark-energy equation of state that an FLRW observer would deduce from observations in a flat $\Lambda$CDM universe endowed with a local inhomogeneity coming from a post-inflationary gaussian density field.}
\label{fig:wofzEx}
\end{figure}

The result of this analysis gives the variance in $w(z)$ at the nodes $z_j=0.05 + 0.1 j$ for $j=1..12$ induced by structures falling in the redshift bin $0.1(i-1) < z_b \le 0.1i$, which is shown in Fig.~\ref{fig:matrix}. 
In Fig.~\ref{fig:wofzEx} we show examples of typical $w(z)$ evolutions as seen by the FLRW observer, affected each by one structure of an arbitrary size. 
It is interesting to see that structures also affect $w(z)$ at redshifts larger than the size of the structure, because even though at larger redshifts the metric is exactly FLRW, the function $z(r)$ (Eq.~\eqref{eq:geodesics}) does not coincide with its FLRW (background) value owing to the structure at the observer.
This effect is more pronounced for the first bin, because there the contrast is largest.

\section{Fitting the luminosity distance}

However, usually one does not derive a fully general $w(z)$ but fits a parameterized model
to the distance data. Typical examples are a constant $w$ or the linear model used for the DETF FoM
mentioned in the introduction, $w(a) = w_0+(1-a) w_a$. As the impact of the local structure is
strongest at low redshift, the variance of the fitted parameters will depend on the way the weight of distance data depends on redshift. 
Even a hypothetical perfect SN experiment will have a non-flat redshift distribution of SNe, as the volume per redshift goes down at low reshift, and as the observability of SNe goes down at high redshifts. Therefore, even though assuming infinitely many SNe, we can choose  a specific redshift distribution modeled
to resemble the expected supernova distribution of the Dark Energy Survey (DES).
Specifically, we weight the 12 redshift bins in the range $0<z\le1.2$ by using the binned rms scatter $\sigma_{\rm bin}$ of the Simulated DES Hybrid 10-field Survey reported in the third column of Table 14 of \cite{Bernstein:2011zf}.

The basic approach is then as above: We fix again a fiducial cosmology with the same parameters
as in the previous section (WMAP7+LRG) and sample again the parameter space describing the inhomogeneities in separate redshift bins $0.1(i-1) < z_b \le 0.1i$ using Eq.~(\ref{gauprob}) as likelihood. For each inhomogeneity, we then fit the parameterized model to the computed luminosity distance-redshift relation in the inhomogeneity and determine the best-fit parameters 
$\theta^*_{i}$ by minimizing the following $\chi^{2}$:
\begin{equation}
\label{chi2}
\chi^{2}(\theta_{i}) = \sum_{j=1..12} \frac{[m_{\rm hom}(z_{j};\theta_{i}) - m_{\rm inh}(z_{j})]^{2}}{\sigma_{\rm bin}^{2}(z_{j})} \,,
\end{equation}
where $z_j=0.05 + 0.1 j$, $m_{\rm hom}$ and $m_{\rm inh}$ are the distance moduli of the homogeneous $\Lambda$CDM model and of the inhomogeneous model, respectively (with the latter playing the role of mock data in our context), and we marginalize analytically the likelihood $\propto e^{-\chi^2/2}$ over an unknown offset (and therefore over the Hubble constant).
This time we can optionally also vary $\Omega_k$ and $\Omega_m$ as the parameterized model breaks the degeneracies. Note that since we only consider the best-fit parameters, this analysis is insensitive to the absolute magnitude of the error bars $\sigma_{\rm bin}^{2}(z_{j})$. The best-fit parameters remain the same in the limiting case of infinitely many SNe. The error bars $\sigma_{\rm bin}^{2}(z_{j})$ have the purpose of imposing a natural redshift distribution.

\begin{figure}
\includegraphics[width=\columnwidth]{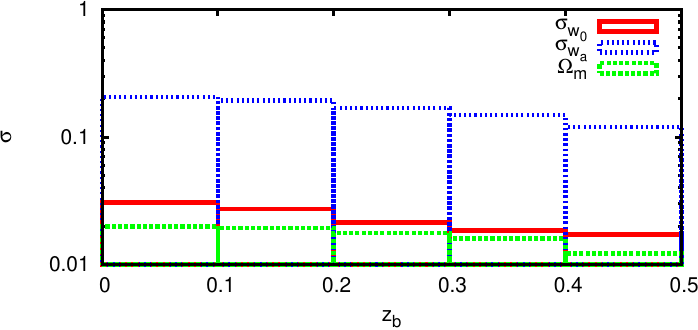}
\caption{Dispersion induced by all structure up to a redshift $z_b$ in bins $0<z_b\le0.1$, $\dots$, $0.4<z_b\le0.5$ on the fit parameters $\{\Omega_m, w_0, w_a\}$ when $\Omega_k = 0$ is assumed. The last column in Table~\ref{tab:var} gives the corresponding values when one averages the effect of these structures.}
\label{fig:fitpars}
\end{figure}

In Fig.~\ref{fig:fitpars} we show the dispersion induced by all structure up to a redshift $z_b\le0.5$ on the fit parameters $\{\Omega_m, w_0, w_a\}$ when the observer assumes that $\Omega_k = 0$. When going to larger redshifts, the modelled effect goes down as $\sigma_L$ decreases for large $L$ (see Fig.~\ref{fig:wofzEx}). If one considers a very large $L$, then all structure up to that $L$ is smoothed over the entire distance $L$. We wish to consider local structure at the observer, which would appear as absent when one smoothes over too large $L$. Therefore we should limit the search to small redshifts. For the same reason, taking an average over different redshifts is only meaningful if the effect does not vary much. Hence it is safe to consider the average variance induced by structures up to $z_b=0.5$. Moreover, for redshifts larger than $z_b=0.5$, other contributions such as lensing start to become important. We therefore compute the expected variance for such structures by combining the MCMC chains for the size bins up to $z_b=0.5$.
 We list in Table \ref{tab:var} the numerical values of the dispersions for the four different parameterized models.
These dispersions 
show that the density perturbations around us, stemming from a standard inflationary spectrum and of unknown density, add an extra uncertainty to these parameters. Unless a measurement of the monopole of local perturbations becomes possible in the future, we can never measure these parameters at a higher accuracy then the listed variances.

\begin{table}[!ht]
\begin{center}
\begin{tabular}{l|cccc}
\hline
& $\{ \Omega_{m}, \Omega_{k} \}$ & $\{\Omega_{m}, w_{0} \}$ &  $\{\Omega_{m}, \Omega_{k}, w_{0} \}$ & $\{\Omega_{m}, w_{0}, w_{a} \}$  \\\hline
$\sigma_{\Omega_m}$ &  0.013 & 0.0043 &  0.019 &  0.018 \\
$\sigma_{\Omega_k}$ &  0.029 & -- &  0.067 & -- \\
$\sigma_{w_0}$ & -- &   0.020 &  0.057 &  0.025 \\
$\sigma_{w_a}$ & -- & -- & -- &   0.18 
                  \end{tabular}
\end{center}
\caption{Intrinsic $1\sigma$ uncertainties on fitted parameters for four different FLRW models.
In each model all remaining cosmological parameters are fixed to the WMAP7+LRG best-fit cosmology. The free parameteres are fitted to the inhomogeneous distance by minimizing the $\chi^{2}$ of Eq.~(\ref{chi2}). The listed dispersions are the standard deviation of the posterior distribution on the given parameter, marginalized over the other free parameters.}
\label{tab:var}
\end{table}

Regarding the FLRW model where $\{\Omega_{m}, w_{0}, w_{a} \}$ were left free, we observe that -- based on these results -- an experiment like DES can never determine $w_0$ to a precision better than $2.5 \%$, and $w_a$ better than $18 \%$.
The Figure of Merit is defined as FoM$= 1/A$ where $A$ is the area bounded by the 95\% c.l.~contour on the $\{w_{0}, w_{a} \}$ plane, marginalized over $\Omega_m$.
The covariance matrix for $\{w_{0}, w_{a} \}$ (which is the inverse of the Fisher matrix) for this intrinsic noise on the dark-energy equation of state is:
\begin{equation}
C_{\rm min}=\left(
\begin{array}{cc}
 \sigma_{w_{0}}^{2} & \rho \, \sigma_{w_{0}}  \sigma_{w_{a}} \\
\rho \, \sigma_{w_{0}}  \sigma_{w_{a}} &  \sigma_{w_{a}}^{2}
\end{array}
\right) ,
\end{equation}
where $\sigma_{w_{0}}$ and $\sigma_{w_{a}}$ are given in the last column of Table \ref{tab:var} and the correlation is $\rho=-0.924$.
The area is then $A_{\rm min}=\pi \sqrt{\det C_{\rm min}} \Delta \chi^{2}= \pi  \sigma_{w_{0}} \sigma_{w_{a}} \sqrt{1- \rho^{2}} \Delta \chi^{2}$ where $\Delta \chi^{2} \simeq 5.99$ for a 95\% c.l.\ contour. We find $1/\sqrt{\det C_{\rm min}} = 581$ and FoM$_{\rm max}=31$. For reference, for example the Dark Energy Survey (DES) expects to achieve $1/\sqrt{\det C_{\rm min}} = 200 \sim 230$~\cite{Bernstein:2011zf}. Future missions like Large Synoptic Survey Telescope (LSST) are expected to improve on the DES FoM by a factor of 5 -- 10~\cite{Abate:2012za}, which is beyond the limit of precision possible that we have found here.

\section{Conclusions}

We have estimated in a conservative way the intrinsic uncertainty in the reconstruction of the dark-energy equation of state by means of distance measurements. Although this appears naively like a noise that could be averaged out, we observe only one universe, and so this uncertainty will show up as a bias in distance measurements.
This phenomenon is usually referred to as {\em cosmic variance}, and can impact also other observables such as the local Hubble constant~\cite{Marra:2013rba}.
We propose that the scientific community use the results of this paper so as to include this extra, systematic source of error in their analysis. In particular, we give the covariance matrix $C_{\rm min}$ for the linear model $w(a) = w_0+(1-a) w_a$ used for the DETF FoM, which can be easily convolved with any other posterior distribution constraining the parameters $w_0$ and $w_a$. Since the large-scale structure limits strongly the power of distance measurements as a probe of the nature of the dark energy and of the curvature of the universe, one may need to use data e.g.\ from galaxy surveys, weak lensing measurements or from the integrated Sachs-Wolfe effect to reduce its impact at least partially.

In Ref.~\cite{Sinclair:2010sb}, it was proposed to exclude sources at redshifts lower than $z=0.035$ rather than the usual $z=0.02$ in cosmological analyses, in order to minimize the effect of our local universe on the inference of the Dark-Energy density. Here we find that for the time dependence of Dark Energy, a cutoff of an order of magnitude larger is still not sufficient, and hence our ignorance should be included in the error budget.

This analysis can be extended in a number of ways.
For example, one may model the local inhomogeneity using non-spherically symmetric and non-compensated density profiles. 
Even more important would be the inclusion of non-local structures like superclusters and filaments which may increase the noise through lensing. According to results from second-order perturbation theory \cite{BenDayan:2012ct,BenDayan:2013gc} there is a scatter of about
$\sigma_\mu \approx 0.02$ or more in the luminosity distance even for $z > 0.4$ which can lead to a percent-level variation for $w$ also at high
redshifts.

\noindent\paragraph{\bf\emph{Acknowledgements}} 
It is a pleasure to thank
L.~Amendola, B.~Bassett, R.~Durrer, R.~Maartens, G.~Marozzi, T.~Riotto and I.~Sawicki
for useful comments and conversations. 
M.K.~acknowledges funding by the Swiss National Science Foundation.
V.M.~and W.V.~acknowledge funding from DFG through the project TRR33 ``The Dark Universe''. 
WV is supported by a Veni research grant from the Netherlands Organization for Scientific Research (NWO).

\bibliographystyle{elsarticle-num}
\bibliography{refs.bib}

\end{document}